\documentclass[11pt]{article}
\usepackage{cite}
\usepackage{amsmath,amsfonts,amssymb}
\usepackage[small,bf,hang]{caption}
\usepackage{slashed}
\usepackage{color}

\usepackage{extarrows}
\usepackage{hyperref}

\usepackage{geometry}
\usepackage{parskip} 

\def\hybrid{
        \topmargin -20pt
        \oddsidemargin 0pt
        \headheight 0pt \headsep 0pt
        \textwidth 6.25in 
        \textheight 9.5in 
        \marginparwidth .875in
        \parskip 5pt plus 1pt \jot = 1.5ex}

\hybrid

 \csname
@addtoreset\endcsname{equation}{section}

\newcommand{\qq}{\quad , \quad}
\newcommand{\Tr}[1]{\text{Tr}\left(#1\right)}

\newcommand{\Sgen}{\mathcal{S}}
\newcommand{\Hgen}{\mathcal{H}}

\def\moth{\mathsurround=0pt}
\newdimen\zo \zo=0pt

\def\tick{\leaders\hrule height 0.5ex depth 0pt \hskip 0.5pt}
\def\upboxfill{$\moth \setbox\zo\hbox{\tick}%
  \hskip 3pt\hbox to 0pt{$\tick$\hss}\hrulefill \hbox to 7.5pt{$\tick$\hss}$}

\def\dtick{\leaders\hrule height .34pt depth 0.5ex \hskip 0.5pt}
\def\downboxfill{$\moth \setbox\zo\hbox{\dtick}%
  \hskip 2pt\hbox to 0pt{$\dtick$\hss}\hrulefill \hbox to 2pt{$\dtick$\hss}$}

\def\bec{\begin{center}}
\def\ec{\end{center}}

\def\qq{\quad\quad}

\def\Tr{{\rm Tr}}

\def\be{\begin{equation}}
\def\ee{\end{equation}}
\def\bea{\begin{eqnarray}}
\def\eea{\end{eqnarray}}
\def\ba{\begin{array}}
\def\ea{\end{array}}

\begin{document}

\begin{titlepage}
\rightline{}
\rightline{January 2021}
\rightline{HU-EP-20/46-RTG} 
\begin{center}
\vskip 1.5cm
 {\Large \bf{   
 String Dualities at Order $\alpha'^{\,3}$}}
\vskip 1.7cm

{\large\bf {Tomas Codina$^\dag$, Olaf Hohm$^\dag$ and Diego Marques$^*$}}
\vskip 1cm

$^\dag$ {\it   Institute for Physics, Humboldt University Berlin,\\
 Zum Gro\ss en Windkanal 6, D-12489 Berlin, Germany}\\
 
\vskip .3cm

$^*$ {\it   Instituto de Astronom\'ia y F\'isica del Espacio, \\
 Casilla de Correo 67 - Suc. 28 (C1428ZAA), Buenos Aires, Argentina}\\
\vskip .1cm

\vskip .4cm

tomas.codina@physik.hu-berlin.de, ohohm@physik.hu-berlin.de, diegomarques@iafe.uba.ar

\vskip .4cm

\end{center}

\bigskip\bigskip
\begin{center} 
\textbf{Abstract}

\end{center} 
\begin{quote}

We compute  the cosmological  reduction of the fourth powers of the Riemann tensor claimed  to arise  
in string theory at order $\alpha'^{\,3}$,  with overall coefficient proportional to $\zeta(3)$, and show that it is compatible with an $O(9,9)$ symmetry.  
This confirms  the general result in string theory, due to Sen, 
that classical string theory with $d$-dimensional translation invariance  admits an 
$O(d,d)$ symmetry to all orders in  $\alpha'$.

\end{quote} 
\vfill
\setcounter{footnote}{0}
\end{titlepage}

\section{Introduction}

String theory continues to be a  most promising framework   for a consistent theory of quantum gravity. 
At low energies string theories  are described by Einstein's theory of general relativity coupled to 
matter fields, which universally include an antisymmetric tensor (B-field) and a scalar (dilaton).  
Intriguingly, however, even classical  string theory modifies general relativity in two important respects: it includes an infinite number 
of higher-derivative corrections governed by the (inverse) string tension $\alpha'$ \cite{GrossWitten,Gross:1986mw}, and it permits dualities identifying solutions  that are 
drastically different in standard geometry. 
These general features suggest promising scenarios for cosmology \cite{Brandenberger:1988aj,Tseytlin:1991xk,Veneziano:1991ek,Meissner:1991zj,Meissner:1991ge},  
but an immediate obstacle is that the explicit form of the $\alpha'$ corrections is at best  known to the first few orders. 
For Type II string theories not even the first non-trivial higher-derivative corrections, which  arise  at order $\alpha'^{\,3}$, are known completely. 
Moreover, 
notwithstanding  early important work in \cite{Bergshoeff:1995cg,Meissner:1996sa,Kaloper:1997ux}, the compatibility of $\alpha'$ corrections with string dualities 
such as T-duality 
has only in recent years become the focus of attention. 
The `space of duality invariant cosmologies' has been explored to all orders in $\alpha'$ and shown to permit 
novel features \cite{Hohm:2019jgu,Hohm:2015doa}, but it is not known which points in this theory space actual string theories inhabit. 
It is thus a matter of some urgency to find efficient methods  to deal with $\alpha'$ corrections.

In this letter we investigate  T-duality  at order $\alpha'^{\,3}$. Our goal is to analyze whether  the eight-derivative corrections (quartic in the Riemann tensor)
with  overall coefficient proportional to the transcendental $\zeta(3)$  
are compatible with the  
$O(9,9)$ T-duality invariance  upon reduction  to one dimension (cosmic time). 
An important  issue  
is that one should allow for the $O(d,d)$ transformations themselves to receive $\alpha'$ corrections \cite{Tseytlin:1991wr}. 
It indeed follows from Meissner's work on the cosmological reduction to first order in $\alpha'$ that the T-duality transformations in terms 
of standard supergravity fields are $\alpha'$-deformed \cite{Meissner:1996sa,Panvel:1992he,Jack:1999av}. 
However, it is possible to find new field variables for which the T-duality transformations take the standard form. 
A general framework was  
developed in \cite{Hohm:2015doa,Hohm:2019jgu} that systematically uses field redefinitions to bring both  the dimensionally reduced action 
and the most general $O(d,d)$ invariant action
to a form that involves  only first order derivatives. 
The claim is that this procedure eliminates 
all ambiguities resulting from 
the freedom to perform integrations by part and to use lower-order equations of motion to modify higher-derivative terms. 
Upon passing to this canonical field basis the $O(d,d)$ invariance, if present, should take the standard form. 
This yields a systematic procedure to test 
the reduced  actions 
for $O(d,d)$ invariance, 
which has been successfully applied 
to first order in $\alpha'$ in cosmological reductions and, more recently, for general torus compactifications \cite{Eloy:2019hnl}.
While the complete higher-derivative corrections at order $\alpha'^{\,3}$ are not known, the eight-derivative terms involving only the metric are 
believed to be known completely. We will see that this is sufficient to show compatibility with $O(d,d)$,  which in turn determines  the 
B-field and dilaton couplings that survive upon cosmological reduction.

Apart from potential applications in cosmology, this result is of conceptual interest in 
view of these eight-derivative corrections being proportional to the transcendental number   $\zeta(3)$. 
The transcendentality implies that these corrections cannot be linked by conventional symmetry transformations to corrections with rational coefficients, 
as present in  bosonic and heterotic string theory starting at  first order in $\alpha'$. It  is sometimes questioned  whether 
the $\zeta(3)$ couplings are compatible  with the continuous $O(d,d,\mathbb{R})$. In a previous version of this letter we arrived at the incorrect conclusion 
that these couplings are not  compatible with $O(d,d,\mathbb{R})$ invariance, but this would have been   in quite serious conflict with basic principles of string theory. 
As  shown by Sen, classical (tree-level) string theory 
truncated to states of zero momentum along $d$ directions admits an $O(d,d,\mathbb{R})$ invariance to all orders in  $\alpha'$ \cite{Sen:1991zi}.  
While the original proof was couched in the language of string field theory the argument only relies on the symmetries of the S-matrix 
of this consistently truncated sector. At tree-level, holomorphic factorization yields two independent manifest $O(d,\mathbb{R})$ symmetries, 
and combining this $O(d,\mathbb{R})\times O(d,\mathbb{R})$ invariance with the $GL(d,\mathbb{R})$ symmetry following from diffeomorphism invariance and 
constant shifts of the B-field implies $O(d,d,\mathbb{R})$ invariance \cite{Hohm:2014sxa}. 
Thus, the tree-level corrections at order $\alpha'^{\,3}$ proportional to $\zeta(3)$ really ought to be consistent with $O(d,d,\mathbb{R})$, 
and indeed they are.

\section{Cosmological Reduction}

We now review the leading corrections in Type II string theory, and compute the minimal form of the one-dimensional effective action obtained after a cosmological reduction. The $\alpha'$ corrections in Type II string theory begin at $\alpha'{}^{3}$. The couplings for the gravitational sector were originally computed from four-point scattering amplitudes \cite{GrossWitten}, and later from the sigma-model $\beta$-function \cite{GrisaruVenZanon, GrisaruZanon, FreemanPopeSohniusStelle}. They take the compact form 
	\begin{equation}\label{Jcoupling}
	J(c) \equiv t_8t_8 R^4 + \frac{c}{8} \epsilon_{10} \epsilon_{10} R^4 \ ,
	\end{equation}
where $c=1$ has been determined in the literature, but here we keep it more general in order to see whether this is fixed  
by duality arguments. 
The first term in (\ref{Jcoupling}) is
\begin{align}
	t_8t_8 R^4 = \ & t^{\mu_1 \dots \mu_8} t_{\nu_1 \dots \nu_8} R_{\mu_1 \mu_2}{}^{ \nu_1 \nu_2} R_{\mu_3 \mu_4}{}^{\nu_3 \nu_4} R_{\mu_5 \mu_6}{}^{ \nu_5 \nu_6} R_{\mu_7 \mu_8}{}^{\nu_7 \nu_8} \nonumber \\
	= \ & 3 \cdot 2^7 \left[ R_{\alpha \beta \mu \nu} R^{\beta \gamma \nu \rho} R^{\sigma \mu}{}_{\gamma \delta} R^{\delta \alpha}{}_{\rho \sigma} + \frac{1}{2} R_{\alpha \beta \mu \nu} R^{\beta \gamma \nu \rho} R_{\gamma \delta \rho \sigma} R^{\delta \alpha \sigma \mu} \right. \nonumber \\
	&- \frac{1}{2} R_{\alpha \beta \mu \nu} R^{\beta \gamma \mu \nu} R_{\gamma \delta \rho \sigma} R^{\delta \alpha \rho \sigma} - \frac{1}{4} R_{\alpha \beta \mu \nu} R^{\beta \gamma \rho \sigma} R^{\mu \nu}{}_{\gamma \delta} R^{\delta \alpha}{}_{\rho \sigma}\nonumber\\
	& \left.+ \frac{1}{16} R_{\alpha \beta \mu \nu} R^{\beta \alpha \rho \sigma} R^{\gamma \delta \mu \nu} R_{\delta \gamma \rho \sigma} + \frac{1}{32} R_{\alpha \beta \mu \nu} R^{\alpha \beta \mu \nu} R_{\gamma \delta \rho \sigma} R^{\gamma \delta \rho \sigma} \right]\ , \label{t8t8}
	\end{align}	
	where the $ t_8 $ tensor  can be  defined by its action over generic matrices  \cite{GreenSchwarz, Schwarz} 
	\begin{eqnarray}
	&&  \!\!\!\!\!\!\!\!\!\!\!\!\!\!\!\!t^{\alpha \beta \gamma \delta \mu \nu \rho \sigma}  M^1_{\alpha \beta} M^2_{\gamma \delta} M^3_{\mu \nu} M^4_{\rho \sigma}=\ 8 \, {\rm Tr} \{ M^1 M^2 M^3 M^4  +   M^1 M^3 M^2 M^4  +   M^1 M^3 M^4 M^2\}  \\
	&& \!\!\!\!\!\!\!\!\!\! -2 \left( {\rm Tr} \{ M^1 M^2\} {\rm Tr}\{M^3 M^4\} + {\rm Tr} \{ M^1 M^3\} {\rm Tr}\{M^2 M^4\} + {\rm Tr} \{ M^1 M^4\} {\rm Tr}\{M^2 M^3\}\right)\ .\nonumber
	\end{eqnarray}	
	 For the second term in (\ref{Jcoupling}) we have the following convention for the Levi-Civita tensor
	\begin{equation}\label{e10e10}
	\begin{aligned}
	\epsilon_{10} \epsilon_{10} R^4 = \ & \epsilon^{\alpha \beta \mu_1 \dots \mu_8} \epsilon_{\alpha \beta  \nu_1 \dots \nu_8} R_{\mu_1 \mu_2}{}^{ \nu_1 \nu_2} R_{\mu_3 \mu_4}{}^{\nu_3 \nu_4} R_{\mu_5 \mu_6}{}^{ \nu_5 \nu_6} R_{\mu_7 \mu_8}{}^{\nu_7 \nu_8} \\ = \ & - 2 \cdot 8! R_{[\alpha \beta}{}^{\alpha \beta} R_{\gamma \delta}{}^{\gamma \delta} R_{\mu \nu}{}^{\mu \nu} R_{\rho \sigma]}{}^{\rho \sigma} 
	\\
	= \ & 3.2^{10} \left[ R^{\alpha \beta}{}_{\gamma \delta} R^{\gamma \nu}{}_{\mu \beta} R^{\sigma \mu}{}_{\alpha \rho} R^{\delta \rho}{}_{\sigma \nu} + R^{\gamma \delta}{}_{\alpha \beta} R^{\mu \nu}{}_{\gamma \delta} R^{\alpha \rho}{}_{\sigma \mu} R^{\sigma \beta}{}_{\nu \rho} \right.\\
	& \ - \frac{1}{2} R_{\alpha \beta \mu \nu} R^{\beta \gamma \nu \rho} R_{\gamma \delta \rho \sigma} R^{\delta \alpha \sigma \mu} + \frac{1}{2} R_{\alpha \beta \mu \nu} R^{\beta \gamma \mu \nu} R_{\gamma \delta \rho \sigma} R^{\delta \alpha \rho \sigma}\\ 
	& \ \left.- \frac{1}{16} R_{\alpha \beta \mu \nu} R^{\beta \alpha \rho \sigma} R^{\gamma \delta \mu \nu} R_{\delta \gamma \rho \sigma} - \frac{1}{32} R_{\alpha \beta \mu \nu} R^{\alpha \beta \mu \nu} R_{\gamma \delta \rho \sigma} R^{\gamma \delta \rho \sigma} + \dots \right] \ ,
	\end{aligned}
	\end{equation}	
	
	where the dots stand for terms containing Ricci tensors and scalars, which can be eliminated by using field redefinitions, at the expense of introducing dilaton couplings that we will ignore at the moment.
	
	The couplings given by $t_8t_8$ have nonzero contribution at four-graviton level \cite{GrossWitten}, while the  $\epsilon_{10}\epsilon_{10}$ interactions have nonzero contributions starting only at five-graviton level \cite{Zumino}. The presence of this term in the tree-level effective action was inferred  by the $\beta$-function approach in \cite{GrisaruVenZanon, GrisaruZanon, FreemanPopeSohniusStelle}, predicting $c = 1$. This prediction was confirmed  in \cite{Garousi:2013tca} through sphere-level scattering amplitudes of five gravitons. The literature also suggests that this value for $c$ is required by supersymmetry \cite{deRoo:1992zp,Peeters:2000qj} and the emergence of T-duality symmetry in a circle compactification \cite{Garousi,Razaghian:2018svg}. For the specific value $c = 1$ it can be shown using Bianchi identities that the corrections are given by only two terms \cite{GrisaruZanon}
	\begin{equation}
	J(1) = - 3 \cdot 2^6 \left[ R^{\alpha \beta \mu \nu} R_{\mu \nu}{}^{\gamma \delta}  R_{\alpha \gamma}{}^{\rho \sigma} R_{\rho \sigma \beta \delta} - 4 R_{\alpha \beta}{}^{\gamma \delta} R_{\delta \mu}{}^{\alpha \nu} R_{\nu \rho}{}^{\beta \sigma} R_{\sigma \gamma}{}^{\mu \rho}\right] \ . \label{J}
	\end{equation}
	The consensus is that these are the unique purely gravitational terms appearing in the leading $\alpha'$ corrections in Type II string theory.

	The simplest  way to test for  $O(d,d)$ invariance is by performing a cosmological reduction in which the $D=10$ dimensional target space splits into a single temporal external direction and $d=9$ internal ones. The fields only depend on time and we use the following ansatz
	\begin{align}
	G_{\mu \nu} = \text{diag}\left( -n^2 , g_{i j} \right) \ , \qq
	\phi= \frac{1}{2}\Phi + \frac{1}{2} \log(\sqrt{g})\ , \qq B_{\mu \nu} = \text{diag}\left(0,b_{i j}\right)\,,
	\end{align}
	where $\mu, \nu$ are $D=10$ indices and $i,j$ are $d=9$ indices. 
	All partial derivatives but $ \partial_0 \Psi = \partial_t \Psi \equiv \dot{\Psi}$ are set to zero. After the reduction, the effective one-dimensional action can be cast in terms of the following quantities
	\begin{equation}
	L^{i}{}_{j} \equiv g^{i k} \dot{g}_{k j} \ , \qq M^{i}{}_{j} \equiv g^{i k} \dot{b}_{k j} \ ,
	\end{equation}
	plus the lapse function $n$, the lower-dimensional dilaton $\Phi$ and  their time derivatives. 
	
	A method to bring  the effective action to a minimal form that makes it systematic to assess its $O(9,9)$ invariance  was introduced in \cite{Hohm:2015doa,Hohm:2019jgu}. 
	The idea is that the lower dimensional equations of motion (where we have gauge fixed $n = 1$ after varying the action)
	\begin{equation}\label{EOM}
	\begin{aligned}
	\dot{L} &= M^2 + \dot \Phi L \ , \\
	\dot{M} &= M L + \dot \Phi M \ , \\
	\ddot{\Phi} &= \frac{1}{2}\left(\dot \Phi^2 + \frac{1}{4}\Tr \left(L^2 - M^2\right)\right) \ ,\\
	\dot \Phi^2 &= \frac{1}{4} \Tr\left(L^2 - M^2\right) \ ,
	\end{aligned}
	\end{equation}
can be combined with integrations by part to remove all higher derivative terms containing dilatons, and also allow one to remove the derivatives from $L$ and $M$, leaving a final minimal form containing only powers of $L$ and $M$. It was then shown which of these interactions can be cast in terms of the generalized metric 
\begin{equation}
	\Sgen \equiv \Hgen \eta^{-1}  = \begin{pmatrix}
	b g^{-1} & g - b g^{-1} b\\
	g^{-1} & - g^{-1} b
	\end{pmatrix} \ ,
	\end{equation}
so as to make the $O(d,d)$ symmetry manifest, if present.  We refer to \cite{Hohm:2015doa,Hohm:2019jgu} for details on this  procedure.

In the two-derivative case, the parent action
	\begin{equation}
	S_0 = \int d^Dx \sqrt{-G} e^{-2 \phi} \left[ R + 4 \left(\nabla \phi\right)^2 -\frac{1}{12} H^2 \right] \ ,
	\end{equation}
	compactifies to an action where the $O(9,9)$ symmetry is manifest  \cite{Meissner:1996sa,Hohm:2015doa,Hohm:2019jgu}
	\begin{equation}\label{dualityCosmicAction}
	S_0 = \int dt \, e^{-\Phi} \left[ -\dot \Phi^2  -\frac{1}{8} \Tr\big(\dot \Sgen^2 \big)\right] \, ,
	\end{equation}
	where we used that $\Tr\big(\dot \Sgen^2 \big) = 2\, \Tr \left(M^2 - L^2 \right)$.

	In the following  we will simplify  the problem by setting  the B-field to zero, 
	which is sufficient in order to display the cosmological effective action in a duality invariant form. In this case, the zeroth order equations of motion 
	 (\ref{EOM}) allow for the redefinitions 
	\begin{equation}\label{EOM_Bzero}
	\dot{L} \rightarrow \dot \Phi L\ , \quad
	\ddot{\Phi} \rightarrow \frac{1}{2}\left(\dot \Phi^2 + \frac{1}{4}\left( L^2 \right) \right)\ , \quad
	\dot \Phi^2 \rightarrow \frac{1}{4} \left( L^2 \right) \ ,
	\end{equation}
	where from now on we will 
	denote the traces of $d\times d$ matrices by parenthesis, i.e., 
	\begin{equation}
	L^i{}_i = \text{Tr}(L) \equiv (L)\ , \quad
	(L^2)^i{}_i = (L^2) \ , \quad \dots
	\end{equation}
	but we will keep the $\text{Tr}$ notation for the duality covariant $2d\times 2d$ matrix $\Sgen$. Indices are raised and lowered with $g$, namely 	$L_{i j} = g_{i k} L^{k}{}_{j}=\dot{g}_{i j}$.

	In the simplified case with vanishing two-form, the generalized metric is related to $L$ by (with $M, N$ denoting  doubled internal indices)
	\begin{equation}
	\Sgen_M{}^N = \begin{pmatrix}
	0 & g_{i j}\\
	g^{i j} & 0
	\end{pmatrix}  \ , \quad
	(\dot \Sgen^{2 m})_M{}^N = \begin{pmatrix}
	(-1)^m (L^{2m})_i{}^j & 0\\
	0 &  (-1)^m (L^{2m})^i{}_j
	\end{pmatrix} \ ,
	\end{equation}
	and so 
	\begin{equation}
	\Tr \big( \dot \Sgen^{2 m} \big) = 2 \, (-1)^m  \, (L^{2m})\ , \qq \Tr\big(\dot \Sgen^{2 m - 1}\big) = 0 \,, \quad m \in \mathbb{N} \ . \label{LtoS}
	\end{equation}
	This shows that only traces containing even powers of $L$ can be written in terms of the generalized metric; 
	those involving odd powers do not admit a duality covariant expression.
	
	In this language, the reduced Riemann tensor reads
	\begin{align}
	R_{i j k l} &= \frac{1}{2} L_{i [k} L_{l]j}\ , \qq
	R_{i 0 j 0} = -\frac{1}{2} \dot{L}_{i j}  - \frac{1}{4} (L^2)_{i j} \ . \label{RiemannComponents}
	\end{align}
	The reduced action takes the form
		 \be
	  S = \int dt \,e^{-\Phi} \,\Big( -\dot \Phi^2  -\frac{1}{8} \Tr\big(\dot \Sgen^2\big)  + \alpha'^3\, \frac{\zeta(3)}{3 \cdot 2^{14}}\, J(c)\Big)\,, \label{GenericAction}
	 \ee
	 where the normalization of \cite{Gross:1986mw} is recovered upon setting $\alpha'=1$. 
	Here $J(c)$, which was defined in (\ref{Jcoupling}), is evaluated using  (\ref{RiemannComponents}). 
	We now briefly explain how this computation is performed following  the general procedure of \cite{Hohm:2015doa,Hohm:2019jgu}. 
	Inserting  (\ref{RiemannComponents}) into (\ref{Jcoupling}) one obtains terms  involving traces of products of $L$ and $\dot L$. One may then systematically 
	eliminate all terms that contain $\dot L$ and $(L^2)$ as follows: First, one uses the EOM (\ref{EOM_Bzero}) to replace $\dot L \to \dot \Phi L$, leaving terms involving traces of products of $L$ and powers of $\dot \Phi$. Even powers of $\dot \Phi$ can then be eliminated by use of the third equation in (\ref{EOM_Bzero}),  $\dot \Phi^2 \to {\tiny \frac 1 4} (L^2)$.
	Those containing odd powers of $\dot \Phi$ vanish. (To see this use repeatedly the substitution $\dot \Phi^2 \to {\tiny \frac 1 4} (L^2)$ to arrive at terms with a single $\dot \Phi$,  then integrate it by parts to get 
		 $\int dt e^{-\Phi} \dot{\Phi} X(L) = \int dt e^{-\Phi} \dot{X}(L)$. Now using $\dot L \to \dot \Phi L$ gives the first term back with a different coefficient, thus proving that these terms vanish.)
	At this point we are left with traces of powers of $L$, and we now argue that terms containing $(L^2)$ vanish, 
	\be		\int dt e^{-\Phi}  (L^2)X(L)  
	= 0 \ .\ee
	This is proved as follows. One uses the lapse EOM to replace $(L^2)$ by $4 \dot{\Phi}^2$, after which 
	one integrates by part one $\dot{\Phi}$ factor, 
	using $e^{-\Phi}\dot{\Phi}=-\frac{d}{dt}(e^{-\Phi})$. This creates  terms with $\ddot{\Phi}$ and $\dot{L}$, for which ones uses again the EOMs (\ref{EOM_Bzero}) 
	to write the result in terms of $\dot{\Phi}^2$ and $(L^2)$. Finally, one replaces $\dot{\Phi}^2$ by  $\frac{1}{4}(L^2)$, using the lapse equation. 
	This reproduces the original integral, but with a different coefficient, hence proving that it is zero.

	The upshot of the above  procedure is that we may ignore from the beginning  all $\dot{L}$ terms, setting  $R_{i 0 j 0} =  - \frac{1}{4} (L^2)_{i j}$, 
	and then eliminate  all $(L^2)$ contributions at the end of the computation. 
	For the two contributions to $J(c)$ one then  finds
	\be
	\begin{split}
	t_8 t_8 R^4 &\simeq \frac 9 4 (L^8) + \frac{51}{16} (L^4)^2 - 6 (L^3)(L^5) \ ,\\
	e_{10} e_{10} R^4 &\simeq - 90 (L^8) + \frac{45}{2} (L^4)^2 + 48 (L^3)(L^5) \ ,
	\end{split}
	\ee
	where the symbol $\simeq$ indicates that these equalities hold up to EOMs and integration by parts inside the integral $\int dt e^{-\Phi} $. 
	We then find that the reduction of (\ref{Jcoupling}) is given by
	\begin{equation}
	J(c) \simeq \frac{1}{4} (9 - 45 c) (L^{8}) + \frac{1}{16} (51 + 45 c) (L^{4})^2 -  6 (1 -   c)(L^{3})(L^{5}) \ .
	\end{equation}
	As explained, while the first two terms can be written in a $O(9,9)$ invariant form using the identities (\ref{LtoS}), the last one involves traces of odd powers of $L$ and is then non-invariant. These contributions come from the second term in the second line of (\ref{t8t8}) and the first term in the fourth line of (\ref{e10e10}), respectively.

	Given that these contributions are unambiguous, meaning that they cannot be modified through EOMs nor integrations by parts, T-duality then  fixes the coefficient to its expected value $c = 1$,
	for which
		\begin{eqnarray}
	J(1)  &\simeq& -9 (L^{8}) +  6 (L^{4})^2 = - \frac 9 2 \Tr (\dot\Sgen^8) + \frac 3 2 \big( \Tr(\dot\Sgen^4)\big)^2 \ .\label{effectiveaction}
	\end{eqnarray}
	
	Using this in (\ref{GenericAction}) we then finally arrive at the minimal form of the cosmological effective action written in a manifestly $O(9,9)$ invariant way
\begin{equation}
	S = \int dt \,e^{-\Phi} \left\{ -\dot \Phi^2  -\frac{1}{8} \Tr \big(\dot {\cal S}^2 \big) + \frac{\alpha'{}^{3} \zeta(3)}{2^{15}}\left[ 
	-3\, \Tr\big(\dot {\cal S}^8\big) +    
	 \Tr \big(\dot {\cal S}^4\big) \Tr \big(\dot {\cal S}^4\big)\right]  \right\}      \ . \label{finaleffectiveaction}
\end{equation}

Let us finally point out that while  the above result was computed for vanishing B-field and dilaton, by duality invariance (\ref{finaleffectiveaction}) must be the 
complete cosmological action including B-field and dilaton. One might be worried that there could be dilaton couplings in higher dimensions that upon cosmological reduction 
and field redefinitions contribute to the gravitational terms and change the above coefficients, but one may convince oneself that this cannot happen. 
Following the steps outlined above  it can be seen that a generic term of the form $\nabla_{\mu_1} \nabla_{\mu_2} \dots \nabla_{\mu_n} \phi\, X^{\mu_1 \mu_2 \dots \mu_n}$, where $X$ is a tensor depending on $G$ and $\phi$, compactifies to terms that can be redefined away, up to terms 
 containing traces $(L)$, which violate duality invariance.

\section{Discussion}

In this letter we have analyzed the compatibility of the continuous $O(d,d,\mathbb{R})$ symmetry with the eight-derivative couplings quartic in the Riemann tensor 
proportional to $\zeta(3)$ that  arise in string theory at order $\alpha'^{\,3}$. 
We have shown that demanding $O(9,9,\mathbb{R})$ invariance upon cosmological reduction to one timelike dimension fixes the relative coefficient between 
the couplings $t_8t_8R^4$ and $\epsilon_{10}\epsilon_{10}R^4$ uniquely to the value determined independently by other methods. 
This result illustrates the strength of duality invariance, for it  allows one 
to reconstruct from the $t_8t_8R^4$ term, which was  computed  by Gross and Witten from the  four-point amplitude \cite{Gross:1986mw}, 
the complete gravitational couplings in ten dimensions, in addition to  determining  the B-field and dilaton couplings that survive 
upon cosmological reduction. This gives strong constraints on the possible B-field and dilaton couplings also in ten dimensions, 
but we do not expect these to be determined completely, since there could be couplings that disappear upon reduction. 

The results presented here are useful, in particular, in that they determine the first two non-trivial coefficients for Type II string theory 
in the general cosmological classification to all orders in $\alpha'$ \cite{Hohm:2019jgu}. 
There have already been a number of papers exploring cosmological consequences of this $\alpha'$-complete cosmology, 
see \cite{Krishnan:2019mkv,Wang:2019kez,Bernardo:2020nol,Bernardo:2020bpa,Nunez:2020hxx}, and here we have  further constrained  the `space of duality covariant string cosmologies'. 
Moreover, our results provide a non-trivial test for the core assumption underlying this classification: that there is a field basis for which 
the $O(d,d,\mathbb{R})$ symmetry, expected to exist in string theory to all orders in $\alpha'$, takes the standard form. 
This is in contrast to double field theory \cite{Hohm:2014xsa} and 
conventional dimensional reduction  with a  generic number of external dimensions \cite{Baron:2017dvb,Eloy:2019hnl}, 
where a Green-Schwarz-type  mechanism needs to be invoked that  can be viewed as $\alpha'$-deforming the $O(d,d)$ transformations.

\textit{Note added:}    Upon completion of  this letter  we became aware of the results in \cite{Wulff}, 
which exhibit obstacles for a double field theory formulation of the $\zeta(3)$ couplings in ten dimensions. 
The first version of this letter also identified an obstacle for the conventional realization of $O(d,d)$ in dimensional reduction, 
but this was due to  a computational mistake, and the corrected results presented here fully confirm 
the presence of $O(d,d)$. We therefore expect that there is also a double field theory formulation of the $\zeta(3)$ couplings, 
perhaps involving novel structures.

\subsection*{Acknowledgements} 

We thank Juan Maldacena, Ashoke Sen and Pierre Vanhove  for discussions and correspondence, 
Stanislav Hronek and Linus Wulff for giving us access to \cite{Wulff} before submission, 
and Heliudson Bernardo, Axel Kleinschmidt, Krzysztof Meissner, Gabriele Veneziano and  Linus Wulff for comments on the first 
version of this letter. 

This work is supported by the ERC Consolidator Grant ``Symmetries \& Cosmology". D.~M.~is supported by CONICET. 
T.~C.~is supported by the Deutsche Forschungsgemeinschaft (DFG, German Research Foundation) - Projektnummer 417533893/GRK2575 ``Rethinking Quantum Field Theory".

\end{document}